\documentclass[onecolumn, 11pt, draftclsnofoot, letterpaper]{IEEEtran}
\ifCLASSINFOpdf
  \usepackage[pdftex]{graphicx}

\else

\fi
\usepackage{amsmath, amssymb, amsfonts, amsthm, subcaption, mathrsfs, balance, xcolor}
\newtheorem{theorem}{Theorem}

\DeclareMathOperator*{\argmax}{\arg\!\max}
\makeatletter
\def\thm@space@setup{\thm@preskip=0pt
\thm@postskip=0pt}
\makeatother

\begin{document}
\title{Fundamental Limits of Spectrum Sharing for NOMA-based Cooperative Relaying}

\author{\IEEEauthorblockN{Vaibhav Kumar\IEEEauthorrefmark{1},
Barry Cardiff\IEEEauthorrefmark{2}, and Mark F. Flanagan\IEEEauthorrefmark{3}}\\
\IEEEauthorblockA{School of Electrical and Electronic Engineering \\
University College Dublin, Belfield, Dublin 4, Ireland\\
Email: \IEEEauthorrefmark{1}vaibhav.kumar@ucdconnect.ie,
\IEEEauthorrefmark{2}barry.cardiff@ucd.ie,
\IEEEauthorrefmark{3}mark.flanagan@ieee.org}}
\maketitle

\begin{abstract}
Non-orthogonal multiple access (NOMA) and spectrum sharing (SS) are two emerging multiple access technologies for efficient spectrum utilization in the fifth-generation (5G) wireless communications standard. In this paper, we present a closed-form analysis of the average achievable sum-rate and outage probability for a NOMA-based cooperative relaying system (CRS) in an underlay spectrum sharing scenario. We consider a peak interference constraint, where the interference inflicted by the secondary (unlicensed) network on the primary-user (licensed) receiver (PU-Rx) should be less than a predetermined threshold. We show that the CRS-NOMA outperforms the CRS with conventional orthogonal multiple access (OMA) for large values of peak interference power at the PU-Rx.
\end{abstract}

\IEEEpeerreviewmaketitle
\section{Introduction}
With the proliferation of wireless communication technologies, services and applications, one of the major challenges for the 5G communication standard is to support large-scale heterogeneous data traffic. NOMA has recently been recognized as a promising multiple access technology for 5G wireless networks as it can accommodate several users within the same orthogonal resource block (time, frequency and/or spreading code) via multiplexing them in the power domain at the transmitter side and using successive interference cancellation (SIC) at the receiver to remove messages intended for other users \cite{HanzoProc}. In the case of NOMA, users with poor channel conditions have a larger share of transmission power, unlike the conventional OMA where more power is allocated to users with strong channel conditions (also known as the water-filling strategy) \cite{PoorApp}. 

Cognitive radio (CR) is another emerging technology intended to enhance the spectrum utilization efficiency in wireless systems via SS.  There are three main SS paradigms: underlay, overlay and interweave \cite{SpectrumGridlock}. In underlay SS, secondary-users (SUs) operate in a frequency band originally owned by a PU such that the interference caused by the SUs on the primary network is less than a predefined limit, often referred to as the \emph{interference temperature}. Therefore, no limit is directly imposed on the power transmitted from a SU transmitter (SU-Tx); it is sufficient that the interference caused at the PU receiver (PU-Rx) is below the threshold. In a fading channel, the secondary network may take advantage of this fact by opportunistically transmitting at a high power level when the interference channel between SU-Tx and PU-Rx is in a deep fade.

The closed-form expression for the relation between the secondary channel capacity and the peak/average interference inflicted on the primary user for different fading distributions were quantified in \cite{Sousa}. The interference from the PU transmitter (PU-Tx) to the SU receiver (SU-Rx), also termed as the primary-to-secondary interference, was not considered in \cite{Sousa} and hence the results derived give an upper-bound on the achievable rate for the secondary network. The average achievable rate for the SS system in the low-power regime considering the primary-to-secondary interference were studied in \cite{Sboui}, for general fading channels where in addition to the interference constraint, a transmit power constraint was also imposed on the SU-Tx.

Different models of spectrum sharing NOMA networks including underlay NOMA, overlay NOMA and cognitive NOMA were discussed in \cite{CogNOMA} and it was shown that cooperative relaying can improve reception reliability. As such, cooperative spectrum sharing NOMA networks have lower outage probability compared to their non-cooperative counterparts. A novel secondary NOMA relay assisted spectrum sharing scheme was proposed in \cite{NOMA-SS}, where first the quality-of-service (QoS) of the PU was guaranteed using maximal-ratio combining (MRC) and then the sum-rate of the SUs was maximized. Another interesting application of NOMA for spatially multiplexed transmission using a cooperative relaying system (CRS-NOMA) to enhance the spectral efficiency was presented in \cite{CRS-NOMA}, where the source was able to deliver two different symbols to the destination in two-time slots with the help of a relay (here Rayleigh fading was considered). The CRS-NOMA can be easily seen to be superior to the conventional OMA relaying system in which a single symbol is delivered to the destination in two time slots. A performance analysis of the CRS-NOMA in Rician fading was presented in \cite{CRS_NOMA_Rician}.

In this paper, we present the achievable sum-rate and outage probability analysis of the CRS-NOMA in an underlay SS scenario, where the power transmitted from the SU-Tx and from the relay are constrained by placing a limit on the peak interference power received at the PU-Rx. For clarity of exposition, no other constraint on the transmit power is imposed. While in practice, the transmit power from the SU-Tx or the relay is limited by hardware capabilities and other health-related safety considerations, the rates derived in this paper serve as an upper-bound on the capacity of the CRS-NOMA based SS system under a peak interference power constraint. 

\section{System Model}
\begin{figure}[hbht]
\centering 
\includegraphics[scale=0.4]{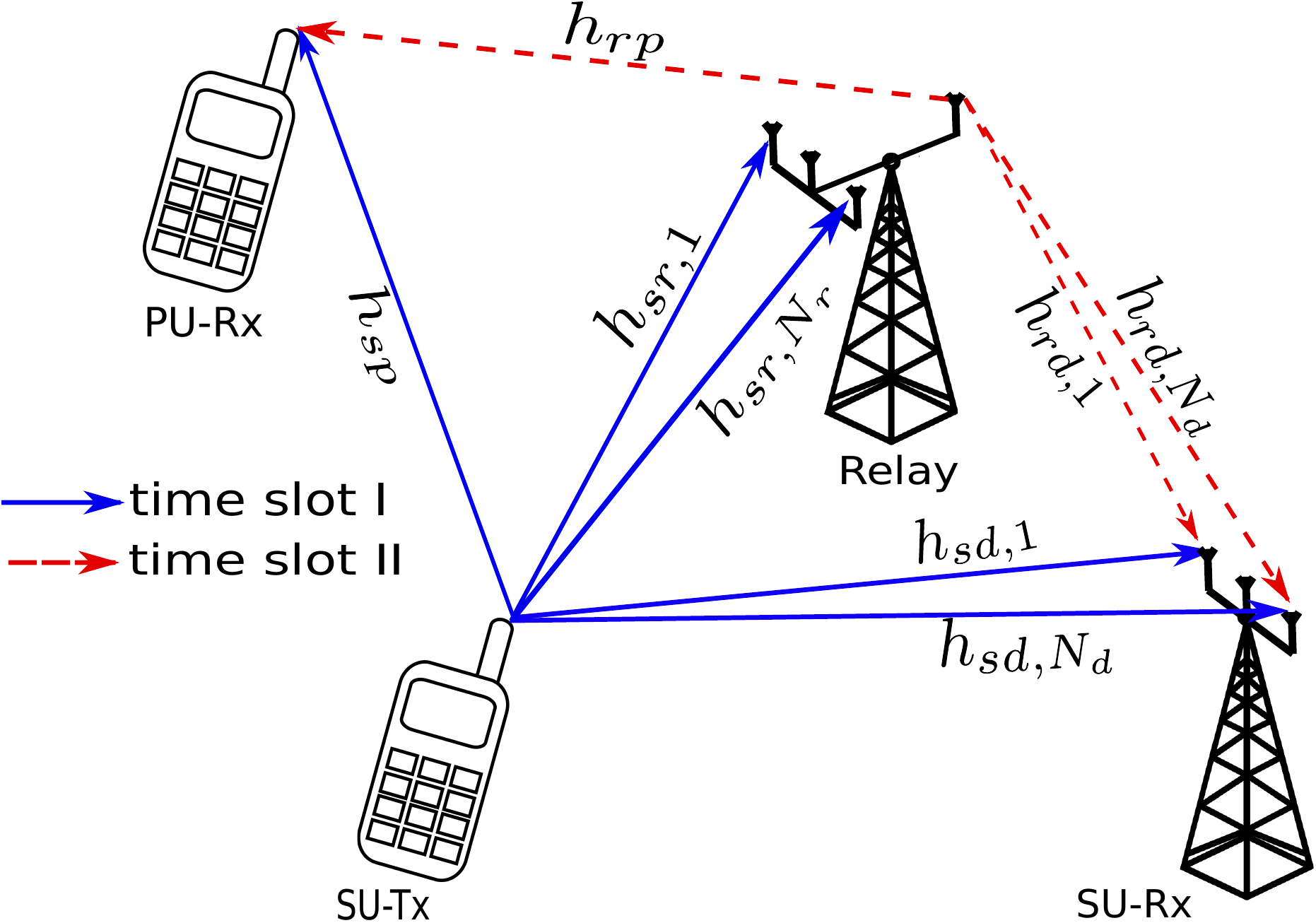}
\caption{System model for CRS-NOMA with underlay spectrum sharing.}
\label{SysMod}
\end{figure}

Consider the spectrum sharing CRS-NOMA shown in~Fig.~\ref{SysMod}, which consists of a SU-Tx $S$, a relay $R$, a SU-Rx $D$ and a PU-Rx $P$. The SU-Tx $S$ is equipped with a single transmit antenna and the PU-Rx $P$ is equipped with a single receive antenna. The relay $R$ is equipped with $N_r$ receive antennas and a single transmit antenna, while the SU-Rx $D$ is equipped with $N_d$ receive antennas. It is assumed that all the nodes are operating in half-duplex mode and all the wireless links are assumed to be independent and Rayleigh distributed. The channel coefficient between the SU-Tx and the $i^{\text{th}}$ receive antenna of the relay $(1 \leq i \leq N_r)$ is denoted by $h_{sr, i}$ and has a mean-square value $\Omega_{sr}$ for all $i$, while that between the SU-Tx and the $j^{\text{th}}$ antenna of the SU-Rx $(1 \leq j \leq N_d)$ is denoted by $h_{sd, j}$ and has a mean-square value $\Omega_{sd}$ for all $j$. 
The channel coefficient between the transmit antenna of the relay and the $j^{\text{th}}$ antenna of the SU-Rx is denoted by $h_{rd, j}$ and has a mean-square value $\Omega_{rd}$ for all $j$. Moreover, the channel coefficient between the SU-Tx and the PU-Rx is denoted by $h_{sp}$ and has a mean-square value $\Omega_{sp}$, while that between the transmit antenna of the relay and the PU-Rx is denoted by $h_{rp}$ and has a mean-square value $\Omega_{rp}$. Furthermore, it is assumed that the channels between the SU-Tx and the SU-Rx are on average weaker that those between the SU-Tx and the relay, i.e., $\Omega_{sd} < \Omega_{sr}$.

In the CRS-NOMA scheme with spectrum sharing, the SU-Tx broadcasts $\sqrt{a_1 P_{s}(h_{sp})} s_1 + \sqrt{a_2 P_s(h_{sp})}s_2$ to both relay and SU-Rx, where $s_1$ and $s_2$ are the data-bearing constellation symbols which are multiplexed in the power domain $(\mathbb E\{|s_i|^2\} = 1$ for $i = 1, 2)$. $P_s(h_{sp})$ is the power transmitted from the SU-Tx and in general is a mapping from the fading coefficient $h_{sp}$ to the set of non-negative real numbers $\mathbb R_{+}$ such that the instantaneous interference at the PU-Rx does not exceed a predetermined value (interference temperature). Moreover, $a_1$ and $a_2$ are power weighting coefficients satisfying the constraints $a_1 + a_2 = 1$ and $a_1 > a_2$. After receiving the signal from the SU-Tx, the SU-Rx decodes symbol $s_1$ treating the interference from $s_2$ as additional noise, while the relay first decodes symbol $s_1$ and then applies SIC to decode $s_2$. In the second time slot, the SU-Tx remains silent and only the relay transmits its estimate of symbol $s_2$, denoted as $\hat{s}_2$, to the SU-Rx with power $P_r(h_{rp})$ which in general is a mapping from the fading coefficient $h_{rp}$ to $\mathbb R_+$ such that the instantaneous interference at the primary receiver does not exceed the predetermined threshold. In this manner, two different symbols are delivered to the secondary receiver in two time slots. 

In contrast to this, in the conventional OMA with underlay spectrum sharing scheme, the SU-Tx broadcasts symbol $s_1$ with power $P_{s}(h_{sp})$ in the first time slot and the relay retransmits the resulting symbol estimate $\hat{s}_1$ to the SU-Rx with power $P_r(h_{rp})$ in the second time slot. The SU-Rx then combines both the copies of symbol $s_1$ and in this manner only a single symbol is delivered to the SU-Rx in two time slots. 

\section{Performance Analysis}
In this section we present the achievable sum-rate and outage probability analysis of the CRS-NOMA with underlay spectrum sharing under peak interference constraint. We first analyze the simple case where $N_r = N_d = 1$ and then we generalize these results to the case when $N_r \geq 1$ and $N_d \geq 1$.
\subsection{Scenario I: $N_r = N_d = 1$}
In this case we denote the channel coefficients for the $S-D$, $S-R$, $R-D$, $S-P$ and $R-P$ links by $h_{sd}$, $h_{sr}$, $h_{rd}$, $h_{sp}$ and $h_{rp}$ respectively. The signals received at the relay, the SU-Rx and the PU-Rx in the first time slot are, respectively, 
\begin{align}
	y_{sr} = & h_{sr} \left( \sqrt{a_1 P_{s}(h_{sp})} s_1 + \sqrt{a_2 P_s(h_{sp})}s_2 \right) + n_{sr}, \notag \\
	y_{sd} = & h_{sd} \left( \sqrt{a_1 P_{s}(h_{sp})} s_1 + \sqrt{a_2 P_s(h_{sp})}s_2 \right) + n_{sd}, \notag \\
	y_{sp} = & h_{sp} \left( \sqrt{a_1 P_{s}(h_{sp})} s_1 + \sqrt{a_2 P_s(h_{sp})}s_2 \right) + n_{sp}, \notag 
\end{align}
where $n_{i}, i \in \{sr, sd, sp\}$ is complex additive white Gaussian noise (AWGN) with zero mean and unit variance. The received instantaneous signal-to-interference-plus-noise ratio (SINR) at the relay for decoding symbol $s_1$ and the received instantaneous signal-to-noise ratio (SNR) for decoding symbol $s_2$ (assuming the symbol $s_1$ is decoded correctly) are given by $\gamma_{sr}^{(1)} = \lambda_{sr} a_1 P_{s}(h_{sp})/\{\lambda_{sr} a_2 P_{s}(h_{sp}) + 1\}$ and  $\gamma_{sr}^{(2)} = \lambda_{sr} a_2 P_{s}(h_{sp})$ respectively, where $\lambda_{sr} = |h_{sr}|^2$. Similarly, the received instantaneous SINR at the SU-Rx for decoding of $s_1$ is given by 	$\gamma_{sd} = \lambda_{sd} a_1 P_{s}(h_{sp})/\{\lambda_{sd} a_2 P_{s}(h_{sp}) + 1\}$, where $\lambda_{sd} = |h_{sd}|^2$. 

In the next time slot, the relay transmits the decoded symbol $\hat{s}_2$ to the SU-Rx with power $P_r(h_{rp})$. The received signals at the SU-Rx and the PU-Rx are, respectively, 
\begin{align}
	y_{rd} = & h_{rd} \sqrt{P_r(h_{rp})} \hat{s}_2 + n_{rd}, \notag \\
	y_{rp} = & h_{rp} \sqrt{P_r(h_{rp})} \hat{s}_2 + n_{rp}, \notag
\end{align}
where $n_{j}, j\in\{rd, rp\}$ is complex AWGN with zero mean and unit variance. The received SNR at the SU-Rx for decoding the symbol $s_2$ is given by $\gamma_{rd} = \lambda_{rd} P_{r}(h_{rp})$, where $\lambda_{rd} = |h_{rd}|^2$. Since the symbol $s_1$ should be decoded correctly at the SU-Rx as well as at the relay for SIC, while satisfying the interference constraint at the PU-Rx, the average achievable rate for symbol $s_1$ is given by~(c.f. \cite[eqn.~(8)]{CRS-NOMA},~\cite[eqn.~(1),~eqn.~(22)]{Sousa})
\begin{align}
	& \bar{C}_{s_1} \!\!=\!\! \max_{P_s(h_{sp}) \geq 0} \int_{|h_{sp}|} \int_{|h_{sr}|} \int_{|h_{sd}|} \!\!\!\!\!\!\!0.5 \log_2 \left( 1 + \min\{\gamma_{sr}^{(1)}, \gamma_{sd}\}\right)  g_{1}(|h_{sp}|) g_{2}(|h_{sr}|) g_{3}(|h_{sd}|)\, d|h_{sp}|\, d|h_{sr}|\, d|h_{sd}|,  \label{Cs1_opt}\\
	& \mathrm{s.t.}\hspace{0.6cm} \lambda_{sp} P_s(h_{sp}) \leq Q,\label{const1}
\end{align}
where $Q$ is the peak interference power that the PU-Rx can tolerate from the secondary network, and where $g_{1}(|h_{sp}|)$, $g_2(|h_{sr}|)$ and $g_3(|h_{sd}|)$ denote the probability density functions (PDFs) of $|h_{sp}|, |h_{sr}|$ and $|h_{sd}|$ respectively. Assuming no other limitation on the power transmitted from the SU-Tx, the optimal transmit power $P_s^*(h_{sp})$ which maximizes\footnote{Here we refer to maximization of the achievable sum-rate for a \emph{given} pair of power allocation coefficients $a_1$ and $a_2$. The achievable sum-rate can be further maximized by optimizing the power allocation coefficients $a_1$ and $a_2$ for a given optimal transmit power level $P_s^* (h_{sp})$.} the achievable rate is given by $Q/\lambda_{sp}$. Hence, the average achievable rate for symbol $s_1$ is given by 
\begin{align}
	 \bar{C}_{s_1} \!\!= & \!0.5 \!\!\int_{\!|h_{sp}|} \!\!\int_{\!|h_{sr}|}\!\! \int_{\!|h_{sd}|} \!\!\!\!\!\!\log_2 \left(1 + \dfrac{\tfrac{\min\{\lambda_{sr}, \lambda_{sd}\}}{\lambda_{sp}}Qa_1}{\tfrac{\min\{\lambda_{sr}, \lambda_{sd}\}}{\lambda_{sp}}Qa_2 + 1} \right)  g_{1}(|h_{sp}|) g_{2}(|h_{sr}|) g_{3}(|h_{sd}|)\, d|h_{sp}|\, d|h_{sr}|\, d|h_{sd}|  \notag \\
	= & \, 0.5 \left[\int_{0}^{\infty} \!\!\!\!\!\!\!\log_2 (\!1 \!+\! Q x\!)f_X\!(\!x\!) dx \!-\!\!\! \int_{0}^{\infty}\!\!\!\!\!\!\!\log_2(\!1 \!+\! Qa_2 x\!)f_X\!(\!x\!) dx\right], \!\!\!\label{Cs1_int}
\end{align}
where $X \triangleq \min\{\lambda_{sr}, \lambda_{sd}\}/\lambda_{sp}$. 
\begin{theorem}
The closed-form expression for the average achievable rate for the symbol $s_1$ is obtained as
\begin{align}
	\bar{C}_{s_1} = & 0.5 \left[ \dfrac{Q \log_2 \left(\frac{Q}{\phi \Omega_{sp}} \right)}{Q - \phi \Omega_{sp}} - \dfrac{a_2 Q \log_2 \left(\frac{a_2 Q}{\phi \Omega_{sp}} \right)}{a_2 Q - \phi \Omega_{sp}}  \right], \label{Cs1_closed}
\end{align}
where $\phi = (1/\Omega_{sr}) + (1/\Omega_{sd})$.
\end{theorem}
\noindent \emph{Proof}: See Appendix A. 

Similarly, the average achievable rate for symbol $s_2$ is given by~(c.f.~\cite[eqn.~(9)]{CRS-NOMA},~\cite[eqn.~(1),~eqn.~(22)]{Sousa})
\begin{align}
	& \bar{C}_{s_2} \!\!=\!\!\!\! \max_{\substack{P_s(h_{sp}) \geq 0 \\ P_r(h_{rp}) \geq 0}} \int_{|h_{sp}|}\!\! \int_{|h_{rp}|} \!\!\int_{|h_{sr}|}\!\!\int_{|h_{rd}|}\!\!\!\!\!\!\!\!\!\!0.5 \log_2\! \left(\! 1\! + \!\min\{\gamma_{sr}^{(2)}, \gamma_{rd}\}\!\right) \!g_1(|h_{sp}|)g_2(|h_{sr}|)g_4(|h_{rp}|)g_5(|h_{rd}|) \notag \\
	& \hspace{11cm} \times d|h_{sp}| d|h_{sr}|d|h_{rp}| d|h_{rd}|, \label{Cs2_opt}\\
	& \mathrm{s.t.} \hspace{0.5cm} \lambda_{sp}P_s(h_{sp}) \leq Q, \label{const2}\\
	&  \hspace{1cm} \lambda_{rp}P_r(h_{rp}) \leq Q, \label{cosnt3}
\end{align}
where $g_4(|h_{rp}|)$ and $g_5(|h_{rd}|)$ denote the PDFs of $|h_{rp}|$ and $|h_{rd}|$ respectively. Assuming no other limitation on the power transmitted from the SU-Tx and the relay, the optimal transmit power levels $P_s^*(h_{sp})$ and $P_r^*(h_{rp})$ which maximize the achievable rate are given by $Q/\lambda_{sp}$ and $Q/\lambda_{rp}$, respectively. Therefore, the average achievable rate for symbol $s_2$ is given by 
\begin{align}
	\bar{C}_{s_2} \!\!= & \!\! \int_{|h_{sp}|}\!\! \int_{|h_{rp}|} \!\!\int_{|h_{sr}|}\!\!\int_{|h_{rd}|}\!\!\!\!\!\!\!\!\!0.5 \log_2\! \left(\! 1\! + \!\min\!\left\{\!\dfrac{\lambda_{sr}a_2}{\lambda_{sp}}, \!\dfrac{\lambda_{rd}}{\lambda_{rp}}\!\right\} \!Q\!\right) \!g_1(|h_{sp}|)g_2(|h_{sr}|)g_4(|h_{rp}|)g_5(|h_{rd}|) \notag \\
	& \hspace{10cm}\times d|h_{sp}| d|h_{sr}| d|h_{rp}| \, d|h_{rd}| \notag \\
	= & \, \!0.5 \!\!\!\int_{0}^{\infty}\!\!\!\!\!\!\! \log_2(\!1\! +\! Qx\!)f_{Y}\!(\!x\!) dx\! = \!\dfrac{1}{2} \log_2(e) Q\!\! \int_{0}^{\infty}\!\!\dfrac{1\! -\! F_{Y}(x)}{1 + Qx}dx, \!\!\!\!\label{Cs2_int}
\end{align}
where $Y \triangleq \min\left\{\lambda_{sr}a_2/\lambda_{sp}, \lambda_{rd}/\lambda_{rp}\right\}$.
\begin{theorem}
The closed-form expression for the average achievable rate for the symbol $s_2$ is given by
\begin{align}
	\bar{C}_{s_2} & = \!\dfrac{0.5 \, a_2 \, \Omega_{rd} \, \Omega_{sr}\, Q }{(\Omega_{rd} \Omega_{sp} \!- \!a_2 \Omega_{rp} \Omega_{sr}) (\Omega_{rd} Q \!- \!\Omega_{rp}) (\Omega_{sp} \!- \!a_2 \Omega_{sr}Q)} \notag \\
	& \hspace{0.5cm}\times \left[ \Omega_{rp} \Omega_{sp} \log_2 \left( \dfrac{a_2 \Omega_{rp} \Omega_{sr}}{\Omega_{rd} \Omega_{sp}}\right) + a_2 \Omega_{rp}\Omega_{sr}Q  \log_2 \left( \dfrac{\Omega_{rd}Q}{\Omega_{rp}}\right) -  \Omega_{rd} \Omega_{sp} Q \log_2 \left( \dfrac{a_2 \Omega_{sr}Q}{\Omega_{sp}}\right)\right]. \label{Cs2_closed}
\end{align}
\end{theorem}
\noindent \emph{Proof}: See Appendix B.

Using \eqref{Cs1_closed} and \eqref{Cs2_closed}, the average achievable sum-rate for the CRS-NOMA is given as 
\begin{align}
	\bar{C}_{\mathrm{sum}} = \bar{C}_{s_1} + \bar{C}_{s_2}. \label{CSum}
\end{align}
For the case of CRS-OMA, the signals received at the relay, the SU-Rx and the PU-Rx in the first time slot are, respectively,
\begin{align}
	y_{sr, \mathrm{OMA}} = & h_{sr} \sqrt{P_s(h_{sr})} s_1 + n_{sr}, \notag \\
	y_{sd, \mathrm{OMA}} = & h_{sd} \sqrt{P_s(h_{sd})} s_1 + n_{sd}, \notag \\
	y_{sp, \mathrm{OMA}} = & h_{sp} \sqrt{P_s(h_{sp})} s_1 + n_{sp}. \notag
\end{align}
In the second time slot, the relay transmits its estimate of $s_1$, denoted by $\hat{s}_1$ to the SU-Rx. The signals received at the SU-Rx and the PU-Rx in the second time slot are respectively,
\begin{align}
	y_{rd, \mathrm{OMA}} = & h_{rd} \sqrt{P_r(h_{rd})} \hat{s_1} + n_{rd}, \notag \\
	y_{rp, \mathrm{OMA}} = & h_{rp} \sqrt{P_r(h_{rp})} \hat{s_1} + n_{rp}. \notag
\end{align}
Following the same peak interference constraint as in the case of NOMA, the average achievable rate for the CRS-OMA system is given as
\begin{align}
	\bar{C}_{\mathrm{OMA}} =  0.5 \mathbb E_Z\left[\log_2 (1 + Q Z)\right], \label{C_OMA}
\end{align}
where\footnote{Here we assume that the SU-Rx applies MRC on the two copies of $s_1$.} $Z \triangleq \min \left\{\tfrac{\lambda_{sr}}{\lambda_{sp}}, \tfrac{\lambda_{sd}}{\lambda_{sp}} + \tfrac{\lambda_{rd}}{\lambda_{rp}}\right\}$ and $E_{\mathcal W}[\cdot]$ denotes the expectation with respect to the random variable $\mathcal W$.
\subsubsection*{Outage probability for CRS-NOMA} We define $\mathcal O_1$ as the outage event for symbol $s_1$, i.e., the event where either the relay or the SU-Rx fails to decode $s_1$ successfully. Hence the outage probability for symbol $s_1$ is given by 
\begin{align}
	\Pr(\!\mathcal O_1\!) \!= & \! \Pr (\!C_{s_1}\!\!\! <\! R_1\!) \!=\! \Pr \!\!\left[\!\dfrac{1}{2}\! \log_2\!\! \left(\!1\! +\! \dfrac{a_1 Q X}{a_2 Q X + 1} \!\right)\! <\! R_1 \!\right] = \Pr(X < \Theta_1)  = \int_{0}^{\Theta_1} \dfrac{\phi \Omega_{sp}}{(1 + \phi \Omega_{sp}x)^2}\, dx \tag{using \eqref{fX}} \\
	= & \dfrac{\phi \Omega_{sp} \Theta_1}{1 + \phi \Omega_{sp}\Theta_1}, \label{s1_outage}
\end{align}
where $C_{s_1}$ is the instantaneous achievable rate for symbol $s_1$, $R_1$ is the target data rate for symbol $s_1$, $\epsilon_1 = 2^{2R_1} - 1$ and $\Theta_1 = \tfrac{\epsilon_1}{Q(a_1 - \epsilon_1 a_2)}$. The integration above is solved using~\cite[eqn.~(3.194-1)]{Grad}. The system design must ensure that $a_1 > \epsilon_1 a_2$, otherwise the outage probability for symbol $s_1$ will always be 1 as noted in~\cite{RelaySelection}. Next, we define $\mathcal O_2$ as the outage event for symbol $s_2$. This outage event can be decomposed as the union of the following disjoint events: (i) symbol $s_1$ cannot be successfully decoded at the relay; (ii) symbol $s_1$ is successfully decoded at the relay, but symbol $s_2$ cannot be successfully decoded at the relay; and (iii) both symbols are successfully decoded at the relay, but symbol $s_2$ cannot be successfully decoded at the SU-Rx. Therefore, the outage probability for the symbol $s_2$ may be expressed as 
\begin{align}
	\Pr(\mathcal O_2) = & \begin{cases} \Pr\left( \!\dfrac{\lambda_{sr}}{\lambda_{sp}}\!< \!\Theta_1\!\right)\! +\! \Pr \left(\! \dfrac{\lambda_{sr}}{\lambda_{sp}}\! \geq\! \Theta_1, \dfrac{\lambda_{sr}}{\lambda_{sp}}\! <\! \Theta_2\!\right)+ \Pr \left( \!\dfrac{\lambda_{sr}}{\lambda_{sp}} \!\geq\! \Theta_2, \dfrac{\lambda_{rd}}{\lambda_{rp}}\! <\! \dfrac{\epsilon_2}{Q}\!\right); & \operatorname{if}\,\,\Theta_1 < \Theta_2 \\
	\Pr \left( \!\dfrac{\lambda_{sr}}{\lambda_{sp}}\! <\! \Theta_1\!\right) + \Pr \left(\! \dfrac{\lambda_{sr}}{\lambda_{sp}} \!\geq\! \Theta_1, \dfrac{\lambda_{rd}}{\lambda_{rp}}\! <\! \dfrac{\epsilon_2}{Q}\!\right); 	& \operatorname{otherwise}
	\end{cases} \notag \\
	= & F_{\scriptscriptstyle{\frac{\lambda_{sr}}{\lambda_{sp}}}}\!(\Theta)\!+\!F_{\scriptscriptstyle {\frac{\lambda_{rd}}{\lambda_{rp}}}}\!\left(\! \dfrac{\epsilon_2}{Q}\!\right)\! -\! F_{\scriptscriptstyle{\frac{\lambda_{sr}}{\lambda_{sp}}}}\!(\Theta)F_{\scriptscriptstyle{\frac{\lambda_{rd}}{\lambda_{rp}}}}\!\!\left(\! \dfrac{\epsilon_2}{Q}\!\right)\!,\label{s2_outage}
\end{align}
where $R_2$ is the target data rate for symbol $s_2$, $\epsilon_2 = 2^{2R_2} -1$, $\Theta_2 = \tfrac{\epsilon_2}{a_2 Q}$ and $\Theta = \max(\Theta_1, \Theta_2)$.

\subsection{Scenario II: $N_r \geq 1$, $N_d \geq 1$}
In this subsection, we generalize the results obtained in the previous subsection for the case when $N_r \!\!\geq\! \!1$, $\!N_d \!\!\geq\!\! 1$ and selection combining (SC), i.e., selection of the antenna with highest instantaneous SNR, is used for reception at both relay and SU-Rx. The received instantaneous SINR at the relay for decoding symbol $s_1$ and the instantaneous SNR for decoding symbol $s_2$ (assuming the symbol $s_1$ is decoded correctly) are given by $\gamma_{sr, \mathrm{SC}}^{(1)} = \delta_{sr} a_1 P_s(h_{sp})/\{\delta_{sr} a_2 P_s(h_{sp}) + 1\}$ and $\gamma_{sr, \mathrm{SC}}^{(2)} = \delta_{sr}a_2 P_s(h_{sp})$, respectively, where $i^* = \argmax_{1 \leq i \leq N_r} |h_{sr, i}|$ and $\delta_{sr} = |h_{sr, i^*}|^2$. Similarly, the received instantaneous SINR at the SU-Rx for decoding symbol $s_1$ is given by $\gamma_{sd, \mathrm{SC}} = \delta_{sd} a_1 P_s(h_{sp})/\{\delta_{sd} a_2 P_s(h_{sp}) + 1\}$, where $j^* = \argmax_{1 \leq j \leq N_d}|h_{sd, j}|$ and $\delta_{sd} = |h_{sd, j^*}|$. In the next time slot, the received instantaneous SNR at the SU-Rx for decoding symbol $s_2$ is given by $\gamma_{rd, \mathrm{SC}} = \delta_{rd} P_r(h_{rp})$, where $k^* = \argmax_{1 \leq k \leq N_d}|h_{rd, k}|$ and $\delta_{rd} = |h_{rd, k^*}|^2$. Following similar arguments as in the previous subsection, the average achievable rate for symbol $s_1$ using SC is given by 
\begin{align}
	\bar{C}_{s_1, \mathrm{SC}} = & 0.5 \int_{0}^{\infty} \log_2 (1 + Q x)f_{\mathcal{X}}(x)\, dx - 0.5 \int_{0}^{\infty} \log_2(1 + Q a_2 x)f_{\mathcal X}(x)\, dx, \label{Cs1_SC_int}
\end{align}
where $\mathcal X \triangleq \min\{\delta_{sr}, \delta_{sd}\}/\lambda_{sp}$.
\begin{theorem}
The closed form expression for the average achievable rate for symbol $s_1$ using SC is obtained as 
\begin{align}
	\bar{C}_{s_1, \mathrm{SC}} = & 0.5 \sum_{k = 1}^{N_r} \sum_{j = 1}^{N_d} (-1)^{k + j} \binom{N_r}{k} \binom{N_d}{j}   \left[ \dfrac{Q \log_2 \!\left( \!\!\frac{Q}{\xi_{k, j} \Omega_{sp}}\!\!\right)\!}{Q - \xi_{k, j} \Omega_{sp}}   - \!\dfrac{a_2 Q \log_2 \left( \frac{a_2 Q}{\xi_{k, j} \Omega_{sp}}\right)}{a_2 Q - \xi_{k, j} \Omega_{sp}} \! \right], \label{Cs1_SC_closed}
\end{align}
where $\xi_{k, j} = (k/\Omega_{sr}) + (j/\Omega_{sd})$.
\end{theorem}
\noindent \emph{Proof}: See Appendix C.

Similarly, the average achievable rate for symbol $s_2$ using SC is given by
\begin{align}
	\bar{C}_{s_2, \mathrm{SC}} = & 0.5 \int_{0}^{\infty} \log_2 (1 + Qx)\, f_{\mathcal Y}(x)\, dx, \label{Cs2_SC_int}
\end{align}
where $\mathcal Y \triangleq \min \left\{ \delta_{sr}a_2/\lambda_{sp}, \delta_{rd}/\lambda_{rp} \right\}$. 
\begin{theorem}
The closed-form expression for the average achievable rate for symbol $s_2$ using SC is obtained as
\begin{align}
	\bar{C}_{s_2, \mathrm{SC}} = & 0.5 Q \left[ \sum_{k = 1}^{N_r} (-1)^{k - 1} \binom{N_r}{k} \dfrac{a_2 \Omega_{sr}  \log_2 \left( \frac{a_2 \Omega_{sr}Q}{k \Omega_{sp}}\right) }{a_2 \Omega_{sr}Q - k \Omega_{sp}} + \sum_{j = 1}^{N_d} (-1)^{j - 1} \binom{N_d}{j} \dfrac{\Omega_{rd} \log_2 \left( \frac{\Omega_{rd}Q}{j \Omega_{rp}}\right) }{\Omega_{rd} Q - j \Omega_{rp}}\right. \notag \\
	& + \!\!\sum_{k = 1}^{N_r} \!\sum_{j = 1}^{N_d} (\!-1\!)^{k + j}\! \binom{N_r}{k}\!\! \binom{N_d}{j} \left\{\!\! \dfrac{k \Omega_{rd}^2 \Omega_{sp}  \log_2 \left( \tfrac{j \Omega_{rp}}{Q\Omega_{rd}}\right)}{(k \Omega_{rd} \Omega_{sp}\! - \!j a_2 \Omega_{rp} \Omega_{sr}) (Q \Omega_{rd}\! - \!j \Omega_{rp})} \right.  \notag \\
	& \hspace{6cm} \left. \left. + \dfrac{j a_2 ^2 \Omega_{rp} \Omega_{sr}^2 \log_2 \left( \tfrac{a_2 Q \Omega_{sr}}{k \Omega_{sp}}\right)}{(k \Omega_{rd} \Omega_{sp}\! - \!j a_2 \Omega_{rp} \Omega_{sr}) (a_2 Q \Omega_{sr}\! - \!k \Omega_{sp})}  \right\} \right]. \label{Cs2_SC_closed}
\end{align}
\end{theorem}
\noindent \emph{Proof}: See Appendix D.

Using \eqref{Cs1_SC_closed} and \eqref{Cs2_SC_closed}, the average achievable rate for the CRS-NOMA using SC is given by 
\begin{equation}
	\bar{C}_{\mathrm{sum, SC}} = \bar{C}_{s_1, \mathrm{SC}} + \bar{C}_{s_2, \mathrm{SC}}. \label{C_sum_SC}
\end{equation}
With some algebraic manipulations, it can be shown that for $N_r = N_d = 1$, \eqref{C_sum_SC} reduces to \eqref{CSum}. For the case of CRS-OMA with SC, the average achievable rate is given as 
\begin{equation}
	\bar{C}_{\mathrm{OMA, SC}} = 0.5 \mathbb E_{\mathcal Z}\left[\log_2(1 + Q \mathcal Z)\right], \label{C_OMA_SC}
\end{equation}
where $\mathcal Z \triangleq \min\{\tfrac{\delta_{sr}}{\lambda_{sp}}, \tfrac{\delta_{sd}}{\lambda_{sp}} + \tfrac{\delta_{rd}}{\lambda_{rp}}\}$.
\subsubsection*{Outage probability for CRS-NOMA with SC} Similar to the previous subsection, we define $\mathscr{O}_1$ as the outage event for symbol $s_1$. Hence the outage probability for symbol $s_1$ using SC is given by 
\begin{align}
	\Pr(\mathscr O_1) = & \Pr \left( C_{s_1, \mathrm{SC}} < R_1\right) = \Pr(\mathcal X < \Theta_1) \notag \\
	= & \int_{0}^{\Theta_1}\!\!\!\!f_{\mathcal X}(x)dx = \sum_{k = 1}^{N_r} \sum_{j = 1}^{N_d} (-1)^{k + j} \binom{N_d}{j}\!\!\binom{N_r}{k}  \dfrac{\xi_{k, j} \Omega_{sp} \Theta_1}{1 + \xi_{k, j} \Omega_{sp} \Theta_1}, \label{s1_SC_outage}
\end{align}
where $C_{s_1, \mathrm{SC}}$ is the instantaneous achievable rate for symbol $s_1$ in the CRS-NOMA using SC. The integration above is solved using~\eqref{f_mathcalX} and~\cite[eqn.~(3.194-1)]{Grad}. Next, we define $\mathscr O_2$ as the outage event for symbol $s_2$ using SC, similar to the previous subsection. Hence, the outage probability for symbol $s_2$ is given by
\begin{align}
	\Pr(\mathscr O_2) = & F_{\scriptscriptstyle{\frac{\delta_{sr}}{\lambda_{sp}}}}\!(\Theta)\!+\!F_{\scriptscriptstyle{\frac{\delta_{rd}}{\lambda_{rp}}}}\!\left( \dfrac{\epsilon_2}{Q}\right)\! -\! F_{\scriptscriptstyle{\frac{\delta_{sr}}{\lambda_{sp}}}}\!(\Theta)F_{\scriptscriptstyle{\frac{\delta_{rd}}{\lambda_{rp}}}}\!\!\left( \dfrac{\epsilon_2}{Q}\right)\!.\label{s2_SC_outage}
\end{align}

\section{Results and Discussion}
\begin{figure}[t]
\centering
\begin{subfigure}{.49\textwidth}
  \centering
  \includegraphics[width=1\linewidth]{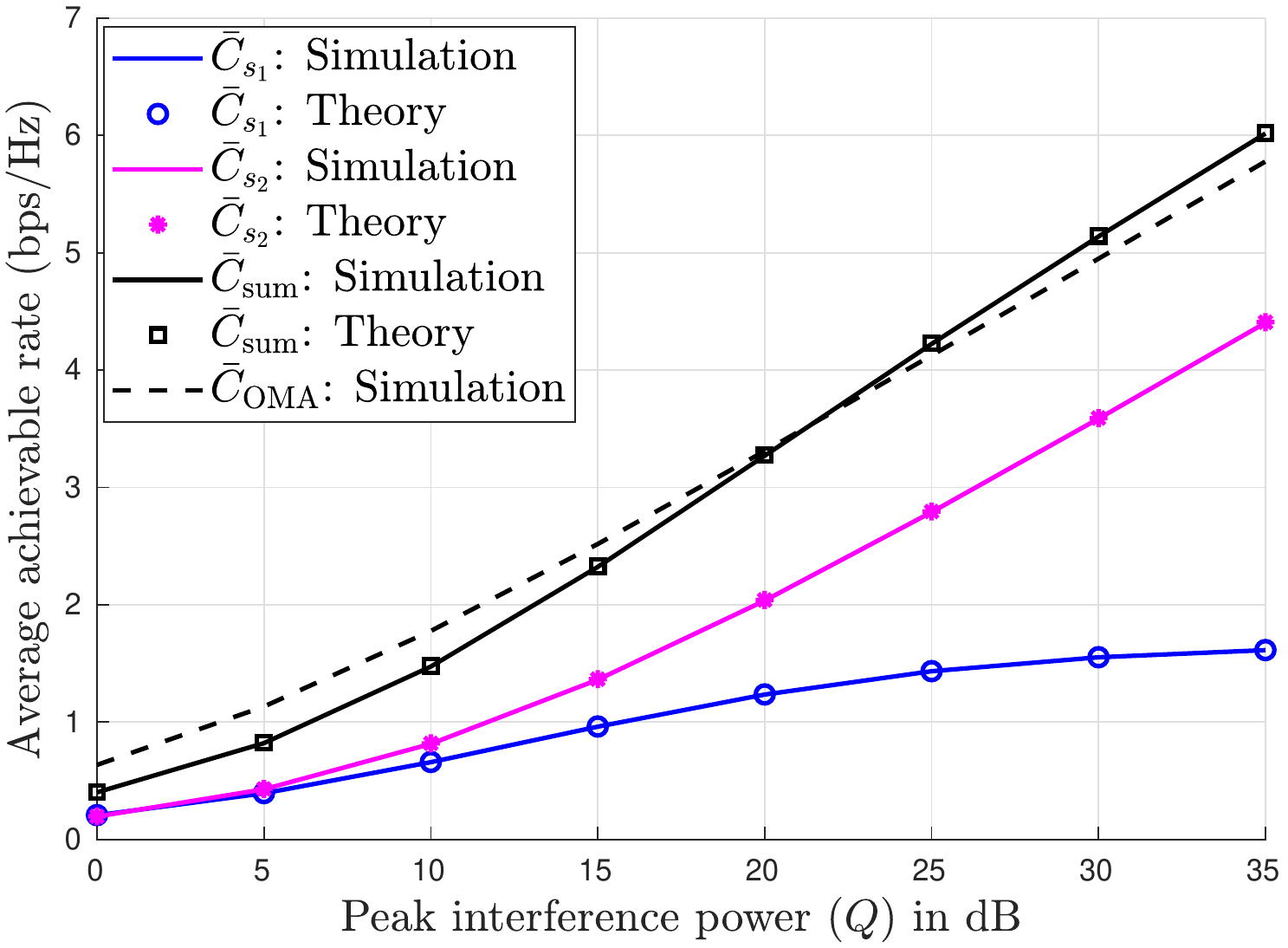}
  \caption{Simple case: $N_r = N_d = 1$.}
  \label{Rate_SingleAntenna}
\end{subfigure}%
\begin{subfigure}{.49\textwidth}
  \centering
  \includegraphics[width=1\linewidth]{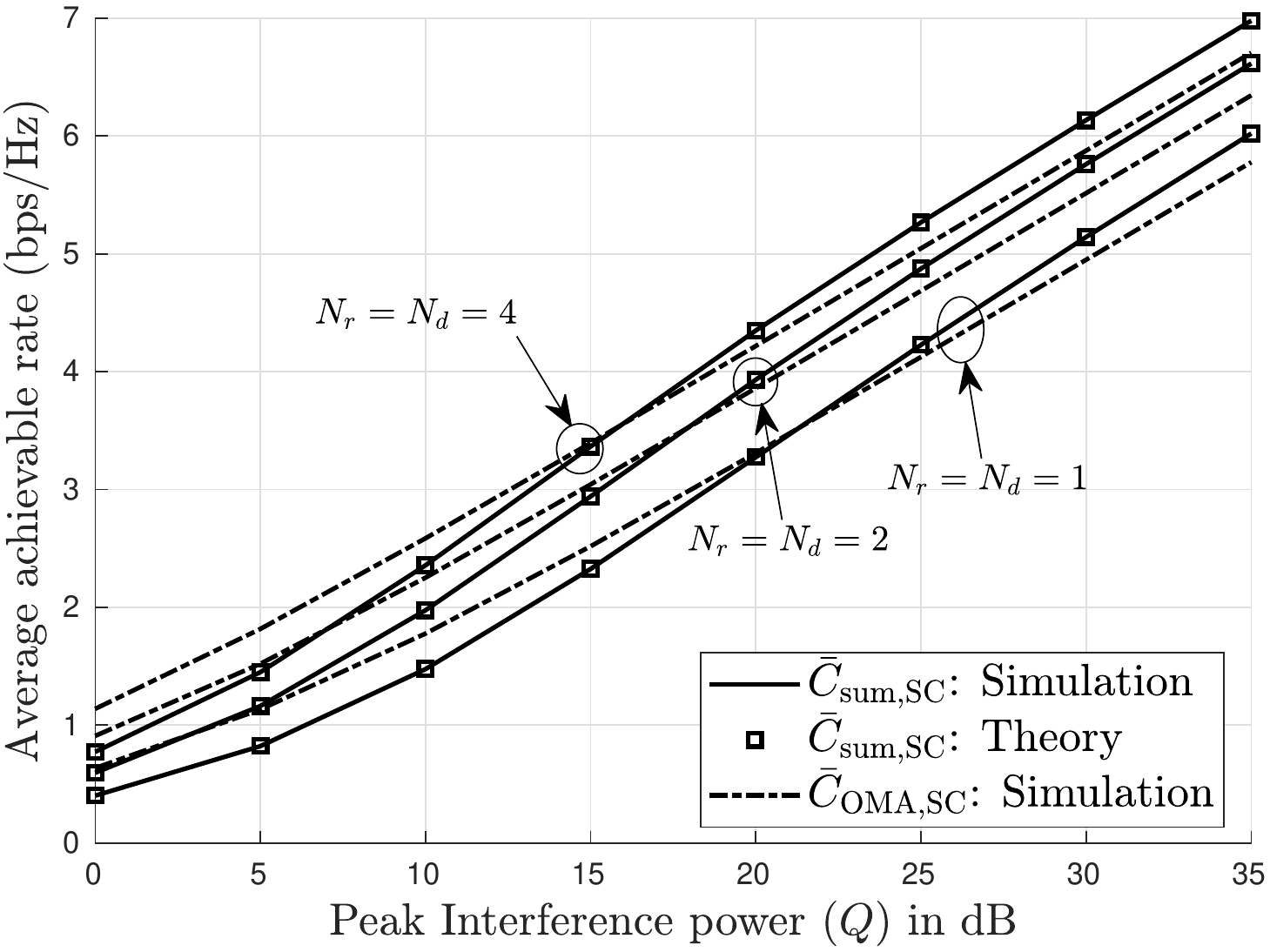}
  \caption{General case: $N_r \!\geq \!1$, $N_d \!\geq\! 1$.}
  \label{Rate_SC}
\end{subfigure}
\caption{Comparison of average achievable rate for CRS based SS.}
\label{Rate}
\end{figure}

In this section, we present the analytical and simulation\footnote{We do not realize the actual scenario for simulation, but rather generate the random variables and then evaluate the average achievable rate.} results for the average achievable rate and outage probability for the spectrum sharing based cooperative relaying system. We consider a scenario where $\Omega_{sd} = 1, \Omega_{sr} = \Omega_{rd} = 10$ and $\Omega_{sp} = \Omega_{rp} = 5.5$. For all the NOMA systems, we consider $a_2 = 0.1, R_1 = R_2 = 1$ bps/Hz. Fig. \ref{Rate_SingleAntenna} shows a comparison of the average achievable rate for the SS based CRS with $N_r = N_d = 1$. It is clear from the figure that for small values of $Q$, the SS based CRS-OMA gives higher achievable rate, but for higher values of $Q$, the CRS-NOMA based SS system outperforms its OMA counterpart and achieves higher spectral efficiency.

In Fig. \ref{Rate_SC}, the average achievable rate for SS based CRS with selection combining is shown for different values of $N_r$ and $N_d$. It is interesting to note that with an increase in the number of receive antennas at the relay and at the SU-Rx, the threshold value $Q$ at which the CRS-NOMA outperforms its OMA counterpart becomes lower. Also, the agreement between the analytical and simulation results in~Fig.~\ref{Rate} confirms the correctness of our sum-rate analysis.

\begin{figure}[t]
\centering
\begin{subfigure}{.49\textwidth}
  \centering
  \includegraphics[width=1\linewidth]{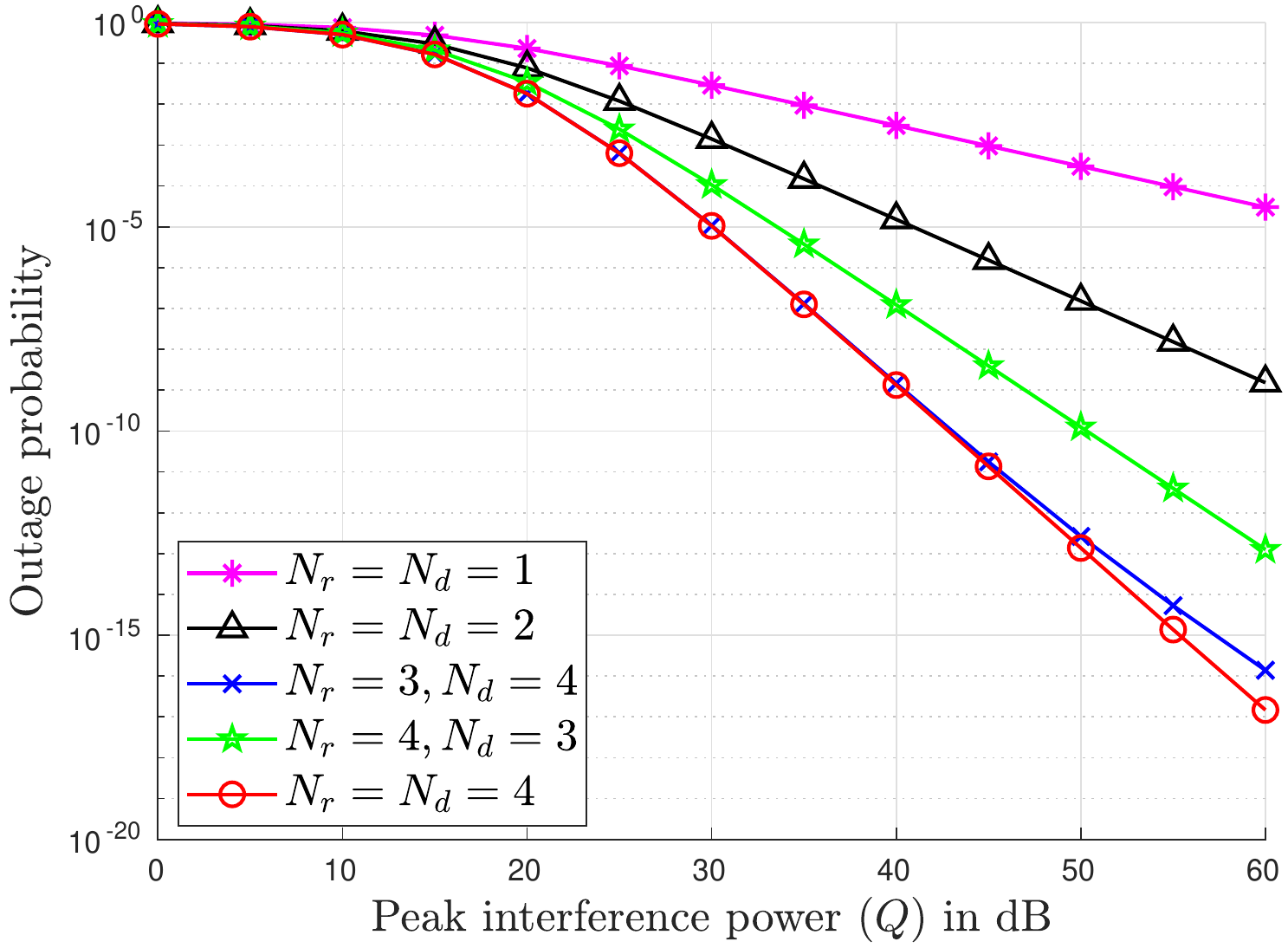}
  \caption{Symbol $s_1$.}
  \label{OutageS1}
\end{subfigure}%
\begin{subfigure}{.49\textwidth}
  \centering
  \includegraphics[width=1\linewidth]{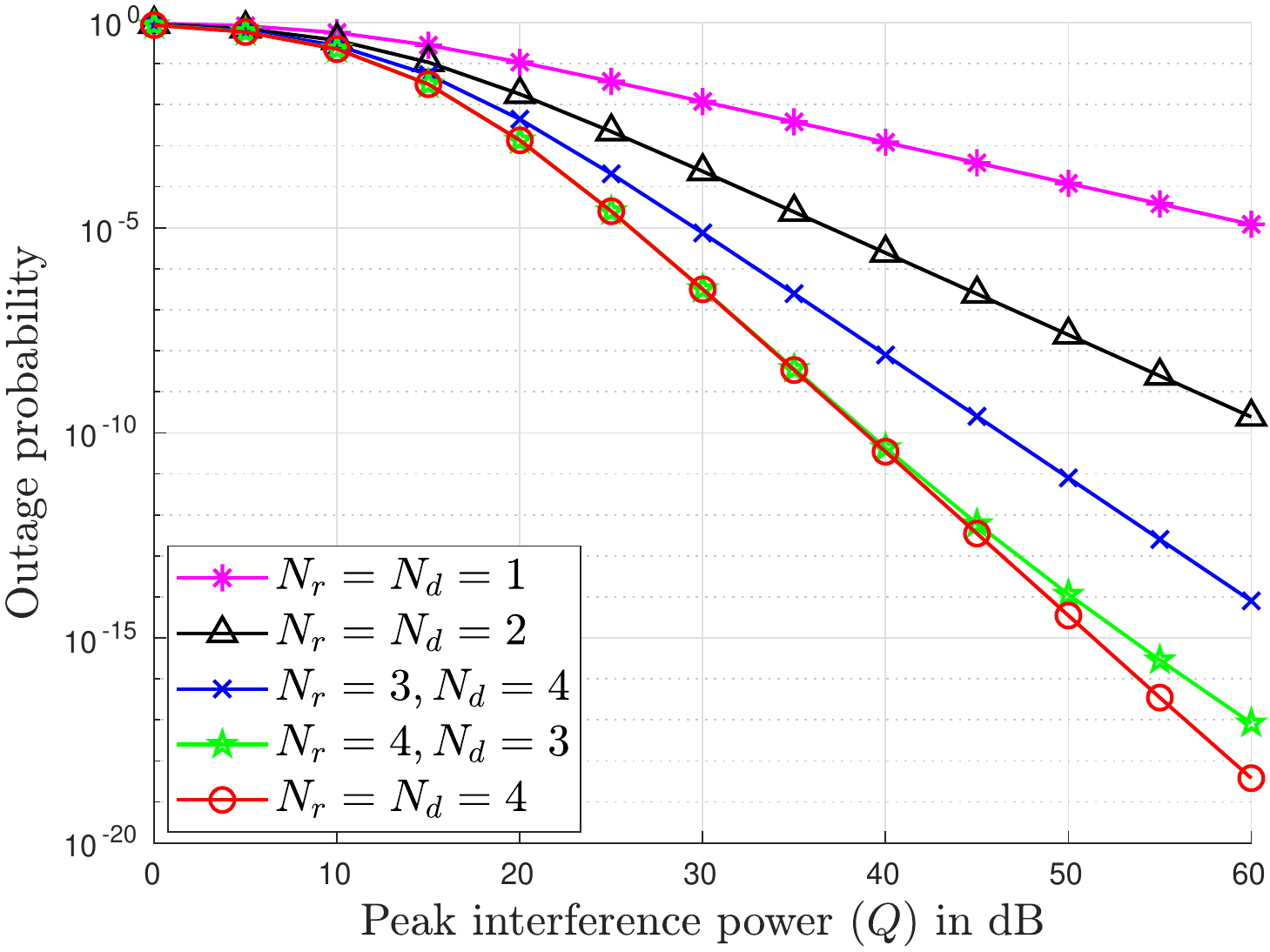}
  \caption{Symbol $s_2$.}
  \label{OutageS2}
\end{subfigure}
\caption{Outage probability for CRS-NOMA with SS.}
\end{figure}

Fig. \ref{OutageS1} and Fig. \ref{OutageS2} show the outage probability for symbol $s_1$ and $s_2$ respectively against $Q$ and as expected, the outage probability decreases as the number of receive antennas is increased at the relay and at the SU-Rx.

\section{Conclusion}
In this paper, we provided a comprehensive achievable sum-rate and outage probability analysis of a NOMA based cooperative relaying system with spectrum sharing considering a peak interference power constraint. We considered the scenario where the relay and the secondary receiver are equipped with multiple receive antennas and where both apply selection combining to combine the received signal. It was shown that for higher values of peak interference power $Q$, the spectrum sharing system based on CRS-NOMA outperforms the spectrum sharing system based on conventional CRS-OMA, achieving higher spectral efficiency. Our results indicate that significant capacity gains can be achieved when the interference channel between the secondary transmitter/relay to primary receiver is in deep fade.
\section*{Acknowledgment}
This publication has emanated from research conducted with the financial support of Science Foundation Ireland (SFI) and is co-funded under the European Regional Development Fund under Grant Number 13/RC/2077.

\appendices
\section{Proof of Theorem 1}
Since $|h_{i}|, i \in \{sr, sd, rd, sp, rp\}$ is Rayleigh distributed, the PDF and the cumulative distribution function (CDF) of $\lambda_i = |h_i|^2$ are, respectively, given by 
\begin{align}
	f_{\lambda_i}(x) = \dfrac{1}{\Omega_i} \exp \left( \dfrac{-x}{\Omega_i}\right),\, F_{\lambda_i}(x) = 1 - \exp \left( \dfrac{-x}{\Omega_i}\right). \notag
\end{align} 
Therefore, the PDF of $\min\{\lambda_{sr}, \lambda_{sd}\}$ is given as\footnote{Given two random variables $\mathcal U$ and $\mathcal V$ with PDFs $f_{\mathcal U}(x)$ and $f_{\mathcal V}(x)$ respectively, and CDFs $F_{\mathcal U}(x)$ and $F_{\mathcal V}(x)$ respectively, the PDF of $\mathcal W \triangleq \min\{\mathcal U, \mathcal V\}$ is given by $f_{\mathcal W}(x) = f_{\mathcal U}(x)[1 - F_{\mathcal V}(x)] + f_{\mathcal V}[1 - F_{\mathcal U}(x)]$ and the CDF of $\mathcal W$ is given by $F_{\mathcal W}(x) = F_{\mathcal U}(x) + F_{\mathcal V}(x) - F_{\mathcal U}(x)F_{\mathcal V}(x)$.}
\begin{align}
	f_{\min\{\lambda_{sr}, \lambda_{sd}\}}(x) = & f_{\lambda_{sr}}(x)[1 \!-\! F_{\lambda_{sd}}(x)]\!+ f_{\lambda_{sd}}(x)[1 \!- \!F_{\lambda_{sr}}(x)] = \phi \exp(-\phi x), \notag 
\end{align}
where $\phi = (1/\Omega_{sr}) + (1/\Omega_{sd})$. Using a transformation of random variables, the PDF of $X = \min\{\lambda_{sr}, \lambda_{sd}\}/\lambda_{sp}$ is therefore given as 
\begin{align}
	& f_{X}(x) = \int_{0}^{\infty} y f_{\min\{\lambda_{sr}, \lambda_{sd}\}}(yx) f_{\lambda_{sp}}(y), dy = \dfrac{\phi}{\Omega_{sp}} \!\int_{0}^{\infty} \!\!\!\!\!\!y \exp \left[ -\left( \!\!\phi x \!+\! \dfrac{1}{\Omega_{sp}}\!\right) y \right]\, dy = \dfrac{\phi \Omega_{sp}}{(1 + \phi \Omega_{sp} x)^2}. \!\!\!\label{fX}
\end{align}
The integral above is solved using~\cite[eqn.~(3.351-3)]{Grad}. Using \eqref{fX}, the first integral in \eqref{Cs1_int} can be solved as 
\begin{align}
	I_1 \!\!\triangleq & \log_2(e) \phi \Omega_{sp}\!\!\int_{0}^{\infty}\!\!\dfrac{\ln (1 + Qx)dx}{(1 + \phi \Omega_{sp} x)^2} \!=\! \dfrac{Q \log_2 \left(\tfrac{Q}{\phi \Omega_{sp}}\right)}{Q - \phi \Omega_{sp}} . \label{I1}
\end{align}
The integration above is solved using~\cite[eqn.~(4.291-17)]{Grad}. Similarly, the second integral in \eqref{Cs1_int} can be solved as 
\begin{align}
	\hspace{-0.2cm}I_2 \!\!\triangleq & \log_2(e) \phi \Omega_{sp}\!\!\int_{0}^{\infty}\!\!\dfrac{\ln (1 + a_2 Qx)dx}{(1 + \phi \Omega_{sp} x)^2}\!\! =\!\! \dfrac{a_2 Q \log_2 \left(\!\tfrac{a_2 Q}{\phi \Omega_{sp}}\!\right)}{a_2 Q - \phi \Omega_{sp}}.\!\! \label{I2}
\end{align}
Hence, using \eqref{Cs1_int}, \eqref{I1} and \eqref{I2}, the closed-form expression for the average achievable rate for symbol $s_1$ in CRS-NOMA reduces to \eqref{Cs1_closed}; this completes the proof.

\section{Proof of Theorem 2}
Using a transformation of random variables, we have 
\begin{align}
	f_{\lambda_{sr}a_2} (x) = \dfrac{1}{a_2}f_{\lambda_{sr}}\left( \dfrac{x}{a_2}\right) = \dfrac{1}{a_2 \Omega_{sr}} \exp \left( \dfrac{-x}{a_2 \Omega_{sr}}\right). \notag 
\end{align}
Hence, 
\begin{align}
	f_{\lambda_{sr}a_2/\lambda_{sp}}(x) = & \int_{0}^{\infty} y f_{\lambda_{sr}a_2}(yx)f_{\lambda_{sp}}(y)\, dy \notag \\
	= & \dfrac{1}{a_2 \Omega_{sr} \Omega_{sp}} \int_{0}^{\infty} y \exp \left[ -\left( \dfrac{x}{a_2 \Omega_{sr}} + \dfrac{1}{\Omega_{sp}}\right) y\right]\, dy = \dfrac{a_2 \Omega_{sr} \Omega_{sp}}{(a_2 \Omega_{sr} + \Omega_{sp}x)^2}, \tag{using \cite[3.351-3]{Grad}} 
\end{align}
and 
\begin{align}
	F_{\lambda_{sr}a_2/\lambda_{sp}}(x) = & \int_{0}^{x}f_{\lambda_{sr}a_2/\lambda_{sp}}(t)\, dt = \dfrac{\Omega_{sp}x}{a_2 \Omega_{sr} + \Omega_{sp}x}. \notag 
\end{align}
The integration above is solved using~\cite[eqn.~(3.194-1)]{Grad}. Similarly, 
\begin{align}
	f_{\lambda_{rd}/\lambda_{rp}}(x) = \dfrac{\Omega_{rd} \Omega_{rp}}{(\Omega_{rd} + \Omega_{rp}x)^2}, \, F_{\lambda_{rd}/\lambda_{rp}}(x) = \dfrac{\Omega_{rp}x}{\Omega_{rd} + \Omega_{rp}x}. \notag 
\end{align}
Therefore, we have 
\begin{align}
	& 1 - F_{Y}(x) \!= \!1 \!-\! F_{\lambda_{sr}a_2/\lambda_{sp}}\!(x) \!-\! F_{\lambda_{rd}/\lambda_{rp}}\!(x) \!+\! F_{\lambda_{rd}/\lambda_{rp}}\!(x) F_{\lambda_{sr}a_2/\lambda_{sp}}(x)  = \dfrac{a_2 \Omega_{sr} \Omega_{rd}}{(a_2 \Omega_{sr} + \Omega_{sp}x)(\Omega_{rd} + \Omega_{rp}x)}. \label{1minusFY}
\end{align}
Using \eqref{Cs2_int} and \eqref{1minusFY}, the average achievable rate for symbol $s_2$ is given as 
\begin{align}
	 & \bar{C}_{s_2} \! = \! \!\! \lim_{\Lambda \to \infty}\!\int_{0}^{\Lambda}\!\!\!\dfrac{0.5 \log_2(e)\,Q \,a_2 \,\Omega_{sr} \Omega_{rd} \, dx}{(a_2 \Omega_{sr} + \Omega_{sp}x)(\Omega_{rd} + \Omega_{rp}x)(1 + Qx)}. \label{Cs2_app}
\end{align}
Solving the integral above using partial fractions, \eqref{Cs2_app} reduces to \eqref{Cs2_closed}; this completes the proof. 

\section{Proof of Theorem 3}
The CDF and PDF of $\delta_{sr}$ are, respectively,
\begin{align}
	F_{\delta_{sr}}(x) = & 1 - \sum_{k = 1}^{N_r} (-1)^{k - 1}\binom{N_r}{k} \exp \left( \dfrac{-kx}{\Omega_{sr}}\right), \notag \\
	f_{\delta_{sr}}(x) = & \sum_{k = 1}^{N_r} (-1)^{k - 1} \binom{N_r}{k} \dfrac{k}{\Omega_{sr}} \exp \left( \dfrac{-kx}{\Omega_{sr}}\right). \notag 
\end{align}
The CDF and PDF of $\delta_{sd}$ can be obtained by replacing $N_r$ by $N_d$ and $\Omega_{sr}$ by $\Omega_{sd}$ in the corresponding equations above. The PDF of $\min\{\delta_{sr}, \delta_{sd}\}$ is given by 
\begin{align}
	f_{\min\{\delta_{sr}, \delta_{sd}\}}\!(x) \!=\!\! \sum_{k = 1}^{N_r} \sum_{j = 1}^{N_d} (\!-1\!)^{k + j} \!\binom{N_r}{k} \!\!\binom{N_d}{j} \xi_{k, j} \exp (\!-\xi_{k, j} x\!), \notag 
\end{align}
where $\xi_{k, j} = (k/\Omega_{sr}) + (j/\Omega_{sd})$. The PDF of $\mathcal X = \min\{\delta_{sr}, \delta_{sd}\}/\lambda_{sp}$ is therefore given by 
\begin{align}
	f_{\mathcal X}(x)\!\! = & \!\! \int_{0}^{\infty}\!\!\!\!\!\!\! y f_{\min\{\delta_{sr}, \delta_{sd}\}}\!(yx)f_{\lambda_{sp}}\!(y)dy = \sum_{k = 1}^{N_r}\!\sum_{j = 1}^{N_d} \! \!\binom{N_r}{k} \binom{N_d}{j}\dfrac{(-1)^{k + j} \xi_{k, j}}{\Omega_{sp}} \!\!\int_{0}^{\infty} \!\!\!\!\!\!y \exp \!\!\left[\! -\!\left( \!\!\xi_{k, j} x\! +\! \dfrac{1}{\Omega_{sp}}\!\!\right) y\right] \, dy \notag \\
	= & \!\sum_{k = 1}^{N_r}\!\sum_{j = 1}^{N_d} \! \!\binom{N_r}{k} \!\!\binom{N_d}{j}\! \dfrac{\!(\!-1\!)^{k + j} \xi_{k, j}}{\Omega_{sp}}\left( \!\!\xi_{k, j} x \!+\! \dfrac{1}{\Omega_{sp}}\!\!\right)^{\!\!-2}.\!\! \label{f_mathcalX}
\end{align}
The integral above is solved using~\cite[eqn.~(3.351-3)]{Grad}. Now, the first integral in \eqref{Cs1_SC_int} can be solved as 
\begin{align}
	 I_3 \triangleq & \log_2(e)\!\!\int_{0}^{\infty}\!\!\!\!\!\!\ln (1\! +\! Qx)f_{\mathcal X}(\!x\!)dx  = \sum_{k = 1}^{N_r}\!\sum_{j = 1}^{N_d} \! \!\binom{N_r}{k} \!\!\binom{N_d}{j} \dfrac{(-1)^{k + j} \xi_{k, j}}{\ln (2)\Omega_{sp}} \!\int_{0}^{\infty} \!\!\!\!\!\!\ln(1 \!+\! Qx) \!\! \left(\! \xi_{k, j} x \!+\! \dfrac{1}{\Omega_{sp}}\!\right)^{-2} \!\!dx \notag \\
	= & \!\sum_{k = 1}^{N_r}\!\sum_{j = 1}^{N_d} \! \!\binom{N_r}{k} \!\!\binom{N_d}{j}  \!\dfrac{\!\!\!\! (\!-1\!)^{k + j} Q}{Q \!-\! \xi_{k, j} \Omega_{sp}}\! \log_2\!\! \left(\! \dfrac{Q}{\xi_{k, j} \Omega_{sp}}\!\right). \label{I3}
\end{align}
The integration above is solved using \cite[eqn.~(4.291-17)]{Grad}. Similarly, the second integral in \eqref{Cs1_SC_int} can be solved as 
\begin{align}
	& I_4 \triangleq  \log_2(e) \int_{0}^{\infty} \ln(1 + Q a_2 x)f_{\mathcal X}(x)\, dx = \sum_{k = 1}^{N_r}\!\sum_{j = 1}^{N_d} \! \!\binom{N_r}{k} \!\!\binom{N_d}{j}  \dfrac{(-1)^{k + j} a_2 Q}{a_2 Q - \xi_{k, j} \Omega_{sp}} \log_2 \left( \dfrac{a_2 Q}{\xi_{k, j} \Omega_{sp}}\right). \label{I4}
\end{align}
Using \eqref{Cs1_SC_int}, \eqref{I3} and \eqref{I4}, the closed-form expression for the average achievable rate for symbol $s_1$ in the CRS-NOMA with SC reduces to \eqref{Cs1_SC_closed}; this completes the proof.

\section{Proof of Theorem 4}
Using a transformation of random variables, the PDF of $\delta_{sr}a_2$ is given as  
\begin{align}
	f_{\delta_{sr}a_2}(x) \!\!=\!\!\dfrac{f_{\delta_{sr}} \left( \tfrac{x}{a_2}\right)}{a_2}  \!=\! \sum_{k = 1}^{N_r} \binom{N_r}{k} \dfrac{(-1)^{k - 1}k}{a_2 \Omega_{sr}} \exp \left( \dfrac{-kx}{a_2 \Omega_{sr}}\right). \notag 
\end{align}
The PDF of $\delta_{sr}a_2/\lambda_{sp}$ is given by 
\begin{align}
	f_{\delta_{sr}a_2/\lambda_{sp}}(x) = & \int_{0}^{\infty} y f_{\delta_{sr}a_2}(yx)f_{\lambda_{sp}}(y)\, dy =\sum_{k = 1}^{N_r}\!\! \binom{N_r}{k} \dfrac{(-1)^{k - 1} k}{a_2 \Omega_{sr} \Omega_{sp}} \!\!\int_{0}^{\infty} \!\!\!\!y \exp \!\!\left[\! -\!\left(\! \dfrac{kx}{a_2 \Omega_{sr}} + \dfrac{1}{\Omega_{sp}}\!\right) y\right]\, dy \notag \\
	= & \sum_{k = 1}^{N_r}\!\! \binom{N_r}{k} \dfrac{(-1)^{k - 1} k}{a_2 \Omega_{sr} \Omega_{sp}} \left(\!\dfrac{kx}{a_2 \Omega_{sr}} + \dfrac{1}{\Omega_{sp}}\! \right)^{-2}. \notag 
\end{align}
The integration above is solved using \cite[3.351-3]{Grad}. Using~\cite[eqn.~(3.194-1)]{Grad}, the CDF of $\delta_{sr}a_2/\lambda_{sp}$ is given by
\begin{align}
F_{\tfrac{\delta_{sr}a_2}{\lambda_{sp}}}(x) = \int_{0}^{x}f_{\tfrac{\delta_{sr}a_2}{\lambda_{sp}}}(t)\, dt = \sum_{k = 1}^{N_r} \binom{N_r}{k} \dfrac{(-1)^{k- 1} k \Omega_{sp}x}{a_2 \Omega_{sr} + k \Omega_{sp}x}. \notag 
\end{align}
Similarly, 
\begin{align}
	f_{\delta_{sr}/\lambda_{rp}}(x) = & \sum_{j = 1}^{N_d}\binom{N_d}{j}\dfrac{(-1)^{j-1} j}{(\Omega_{rd} + \Omega_{rp})} \left( \dfrac{jx}{\Omega_{rd}} + \dfrac{1}{\Omega_{rp}}\right)^{-2}, \notag \\
	F_{\delta_{rd}/\lambda_{rp}}(x) = & \sum_{j = 1}^{N_d}\binom{N_d}{j}\dfrac{(-1)^{j-1}j \Omega_{rp}x}{\Omega_{rd} + j \Omega_{rp}x}. \notag
\end{align}
Therefore, for $\mathcal Y = \min\left\{ \delta_{sr}a_2/\lambda_{sp}, \delta_{rd}/\lambda_{rp}\right\}$, we have 
\begin{align}
	1 - F_{\mathcal Y}(x) = & 1 - F_{\scriptscriptstyle{\frac{\delta_{sr}a_2}{\lambda_{sp}}}}(x) - F_{\scriptscriptstyle{\frac{\delta_{rd}}{\lambda_{rp}}}} + F_{\scriptscriptstyle{\frac{\delta_{sr}a_2}{\lambda_{sp}}}}(x)  F_{\scriptscriptstyle{\frac{\delta_{rd}}{\lambda_{rp}}}}(x) \notag \\
	= & 1 \!-\!\! \sum_{k = 1}^{N_r}\!\! \binom{N_r}{k} \dfrac{(-1)^{k- 1} k \Omega_{sp}x}{a_2 \Omega_{sr} + k \Omega_{sp}x} \!- \!\sum_{j = 1}^{N_d}\binom{N_d}{j}\dfrac{(-1)^{j-1}j \Omega_{rp}x}{\Omega_{rd} + j \Omega_{rp}x} \notag \\
	& \hspace{4.5cm}+ \sum_{k = 1}^{N_r} \!\sum_{j = 1}^{N_d}\! \binom{N_r}{k}\! \binom{N_d}{j}\! \dfrac{(-1)^{k + j} k j \Omega_{sp} \Omega_{rp}x^2}{(a_2 \Omega_{sr} \!+\! k \Omega_{sp}x)(\Omega_{rd} \!+\! j \Omega_{rp}x)}. \label{1minus_FmathcalY}
\end{align}
Using \eqref{Cs2_SC_int}, the average achievable rate for symbol $s_2$ using SC is given as
\begin{align}
	\bar{C}_{s_2, \mathrm{SC}} = & 0.5 \int_{0}^{\infty}\log_2(1 + Qx)f_{\mathcal Y}(x)\, dx = 0.5 \log_2 (e)Q\int_{0}^{\infty}\dfrac{1 - F_{\mathcal Y}(x)}{1 + Qx}\, dx. \label{Cs2_SC_int_alternate}
\end{align}
Now we define the integral $I_5$ as
\begin{align}
	I_5 \triangleq & \int_{0}^{\infty}\dfrac{1}{1 + Q x}\left( \dfrac{k \Omega_{sp} x}{a_2 \Omega_{sr} + k \Omega_{sp}x} \right)\, dx = \int_{0}^{\infty} \dfrac{1}{1 + Q x} \left( 1 - \dfrac{a_2 \Omega_{sr}}{a_2 \Omega_{sr} + k \Omega_{sp}x } \right) \, dx \notag \\
	= & \int_{0}^{\infty} \!\!\!\!\!\!\dfrac{1}{1 + Q x} dx - \!\!\!\lim_{\Lambda \to \infty}\!\int_{0}^{\Lambda} \!\!\!\!\!\dfrac{a_2 \Omega_{sr}\, dx}{(a_2 \Omega_{sr} + k \Omega_{sp}x )(1 + Qx)}\! = \!\!\! \int_{0}^{\infty} \!\!\!\!\!\!\dfrac{dx}{1 + Q x}   \!- \!\dfrac{a_2 \Omega_{sr} \ln \left( \frac{a_2 \Omega_{sr}Q}{k \Omega_{sp}}\right)}{a_2 \Omega_{sr}Q - k \Omega_{sp}}. \label{I5}
\end{align}
The integration above is solved using partial fractions. Similarly, 
\begin{align}
	I_6 \triangleq & \int_{0}^{\infty} \dfrac{1}{1 + Q x} \left( \dfrac{j \Omega_{rp}x}{\Omega_{rd} + j\Omega_{rp}x}\right)\, dx = \int_{0}^{\infty} \dfrac{1}{1 + Qx}\, dx - \dfrac{\Omega_{rd}}{\Omega_{rd}Q - j\Omega_{rp}} \ln \left( \dfrac{\Omega_{rd}Q}{j\Omega_{rp}}\right), \label{I6}
\end{align}
and 
\begin{align}
	I_7 \triangleq & \int_{0}^{\infty} \dfrac{1}{1 + Qx}\left(  \dfrac{k j \Omega_{sp} \Omega_{rp}x^2}{(a_2 \Omega_{sr} + k \Omega_{sp}x) (\Omega_{rd} + j \Omega_{rp}x)}\right) \, dx \notag \\
	= & \int_{0}^{\infty} \!\!\!\!\!\!\!\dfrac{dx}{1 + Qx} \left[ \!1 \!+ \!\dfrac{k \Omega_{rd}^2 \Omega_{sp}}{(a_2 j \Omega_{rp} \Omega_{sr} - k \Omega_{rd} \Omega_{sp}) (\Omega_{rd} + j \Omega_{rp}x)}  + \dfrac{a_2^2 j \Omega_{rp} \Omega_{sr}^2}{(k \Omega_{rd} \Omega_{sp} - a_2 j \Omega_{rp} \Omega_{sr}) (a_2 \Omega_{sr} + k \Omega_{sp}x)}\right] \notag \\
	= & \int_{0}^{\infty} \!\!\!\!\!\!\dfrac{dx}{1 + Qx} \!+\! \dfrac{k \Omega_{rd}^2 \Omega_{sp} \ln \left( \tfrac{j \Omega_{rp}}{\Omega_{rd}Q}\right)}{(k \Omega_{rd} \Omega_{sp} \!-\! a_2 \, j \Omega_{rp} \Omega_{sr})\! (\Omega_{rd} Q \!-\! j \Omega_{rp})}   + \dfrac{a_2 ^2 j \Omega_{rp} \Omega_{sr}^2 \ln \left( \tfrac{a_2 \Omega_{sr}Q}{k \Omega_{sp}}\right)}{(k \Omega_{rd} \Omega_{sp} - a_2 j \Omega_{rp} \Omega_{sr}) (a_2 \Omega_{sr} Q - k \Omega_{sp})} . \label{I7}
\end{align}
Moreover, we also have 
\begin{align}
	& \left[\! 1 \!- \!\sum_{k = 1}^{N_r} (-1)^{k - 1}\! \binom{N_r}{k}\! - \!\sum_{j = 1}^{N_d} (-1)^{j - 1} \binom{N_d}{j} + \sum_{k = 1}^{N_r} \sum_{j = 1}^{N_d} (-1)^{k + j} \binom{N_r}{k} \binom{N_d}{j}\right] \int_{0}^{\infty}\dfrac{dx}{1 + Qx} = 0. \label{Identity}
\end{align}
Using \eqref{Cs2_SC_int_alternate} -- \eqref{Identity}, the closed-form expression for the average achievable rate of symbol $s_2$ in CRS-NOMA using SC reduces to \eqref{Cs2_SC_closed}; this completes the proof.

\bibliographystyle{IEEEtran}
\bibliography{OneColumn_GCWS18}
\end{document}